\documentclass[twocolumn,prl,aps,showpacs,floatfix,superscriptaddress,preprintnumbers,amsmath,amssymb]{revtex4}

\usepackage{graphicx}

\begin{document}

\newcommand{\hto}{Ho$_2$Ti$_{2}$O$_7$}
\newcommand{\dto}{Dy$_2$Ti$_{2}$O$_7$}
\newcommand{\tto}{Tb$_2$Ti$_{2}$O$_7$}

\newcommand{\dmdb}{${\rm d}M/{\rm d}B$}
\newcommand{\dmdbi}{${\rm d}M/{\rm d}B_{\rm int}$}

\newcommand{\dmdh}{${\rm d}M/{\rm d}\mu_0H$}

\title{Vibrating-coil magnetometry of the spin liquid properties of \tto}

\author{S. Legl}
\affiliation{Technische Universit\"at M\"unchen, Physik-Department E21, D-85748 Garching, Germany}

\author{C. Krey}
\affiliation{Technische Universit\"at M\"unchen, Physik-Department E21, D-85748 Garching, Germany}

\author{S. R. Dunsiger}
\affiliation{Technische Universit\"at M\"unchen, Physik-Department E21, D-85748 Garching, Germany}
 
\author{H.A. Dabkowska}
\affiliation{
Brockhouse Institute for Materials Research, McMaster University, Hamilton, Ontario L8S 4M1, Canada
}

\author{J. A. Rodriguez}
\affiliation{
Department of Physics and Astronomy, McMaster University, Hamilton, Ontario L8S 4M1, Canada
}
\affiliation{
 Laboratory for Muon Spin Spectroscopy, Paul Scherrer Institut, 5232 Villigen, Switzerland
}

\author{G.M. Luke}
\affiliation{
Department of Physics and Astronomy, McMaster University, Hamilton, Ontario L8S 4M1, Canada
}

\author{C. Pfleiderer}
\affiliation{Technische Universit\"at M\"unchen, Physik-Department E21, D-85748 Garching, Germany}

\date{\today}

\begin{abstract}
We have explored the spin liquid state in {\tto} with vibrating coil magnetometry down to $\sim0.04\,{\rm K}$ under magnetic fields up to 5\,T. We observe magnetic history dependence below $T^*\sim 0.2\,{\rm K}$ reminiscent of the classical spin ice systems {\hto} and {\dto}. The magnetic phase diagram inferred from the magnetization is essentially isotropic, without evidence of magnetization plateaux as anticipated for so-called quantum spin ice, predicted theoretically for $[ 111]$ when quantum fluctuations renormalize the interactions. Instead, the magnetization for $T\ll T^*$ agrees semi-quantitatively with the predictions of ``all-in/all-out" (AIAO) antiferromagnetism. Taken together this suggests that the spin liquid state in {\tto} is akin to an incipient AIAO-antiferromagnet. 
\end{abstract}


\vskip2pc

\maketitle

%%%%%%%%%%%%%%%%%%%%%%%%%%%%%%%%

% determine correct axis notation
% fix figure 3; zfc and fc/fh; adapt text accordingly
% check ref of L. Balents is in text
% complete ref. 17
% Größe Labels Fig. 4 C recht und links gleich?

Pyrochlore oxides, A$_2$B$_2$O$_7$, in which rare earth magnetic moments  are located on the A-site of a three-dimensional  network of corner sharing tetrahedra are model systems of geometric  frustration~\cite{Gard09}. The consequences of such geometric frustration are intimately connected with the strength and the nature of the magnetic anisotropy at the rare earth site. For instance, in {\hto} and {\dto} a strong easy-axis (Ising)-anisotropy along the local $[111]$ axis in the unit cell, together with net ferromagnetic interactions, are the most important preconditions for the emergence of the highly celebrated spin ice behavior~\cite{Bram02, Ryzh05, Cast08, Ging09}. 

A major unresolved question in geometric frustration concerns the fate of the spin ice state, when the strength of the local Ising anisotropy is reduced. This may boost the relative importance of quantum fluctuations.  In fact, an exciting theoretical proposal states that quantum fluctuations may renormalise the exchange interactions of an unfrustrated $\langle 111 \rangle $ antiferromagnetic Ising antiferromagnet, making them effectively ferromagnetic.  The associated novel state is referred to as quantum spin ice (QSI)~\cite{Mola07,Mola09,Mola09a}, and may be viewed as a spin liquid in which thermal longitudinal spin fluctuations which break the spin ice rules, as well as thermal and quantum fluctuations transverse to the local $\langle 111 \rangle$ directions are relevant. Compelling evidence of a QSI would be magnetization plateaux like those observed in {\hto} and {\dto}, for a magnetic field strictly along a global $\langle111\rangle$ axis~\cite{Saka03,Petr03,Petr11,Krey11}. In turn an anisotropy of the magnetic phase diagram would be expected reminiscent of classical spin ice. 

An ideal model system to study whether any of the classical spin ice properties survive under reduced  local magnetic anisotropy is {\tto}. At high temperatures {\tto} exhibits a Curie-Weiss susceptibility with a large effective moment $\mu_{\rm eff}=9.6\mu_{\rm B}\rm Tb^{-1}$ and a negative Curie-Wei\ss \, temperature $\Theta_{\rm CW}$ characteristic of antiferromagnetic interactions \cite{Gard99,Mire07}. As the crystal electric field (CEF) of the Tb$^{3+}$ ion leads to a ground state doublet and an energy gap of $\sim 18$\,K to the first excited state~\cite{Ging00a,Mire07}, antiferromagnetic order is expected around $\sim1\,{\rm K}$ \cite{Hert00}. 
Surprisingly however, $\mu$SR \cite{Gard99, Duns03}, the ac susceptibility \cite{Hert00,Gard03}, and Neutron Spin Echo (NSE)~\cite{Gard03, Gard04} established strong spin dynamics down to 20\,mK without long-range magnetic order. 

The origin and nature of the lack of long-range magnetic order in {\tto} represents a major puzzle in geometrically frustrated magnetism.  It must however be a sensitive function of the low lying energy levels.   X-ray diffraction \cite{Ruff07} and a related transition at $\sim$0.15\,K~\cite{Yaouanc11} suggest a  cooperative Jahn-Teller distortion. A possible splitting of the doublet to yield a singlet ground state has been inferred from specific heat data~\cite{Chap10}, though this is at odds with high energy resolution neutron-scattering~\cite{Gaulin11}. The large electronic-nuclear hyperfine coupling of Tb suggests that nuclear degrees of freedom may also be important.  It is unresolved if evidence for magnetic glassiness in the milli-kelvin regime is intrinsic \cite{Luo01} or due to defects~\cite{Yasu02,Gard03}. Moreover, for magnetic fields applied along $\langle110\rangle$ the specific heat~\cite{Hama04} and neutron scattering~\cite{Rule06} suggest field-induced order with spin-ice like properties~\cite{Cao08, Guka09, sazo10}, while the spin order for field along $\langle111\rangle$ is undetermined~\cite{Yasui01}.  Magneto-elastic effects are also known to be very pronounced~\cite{Ruff07,Aleks85,Mamsu86}, consistent with the observation of pressure induced antiferromagnetic order \cite{Mire07}. 

Calculations exploring the role of the interaction strength and CEF splitting suggest that {\tto} is at the border between a QSI and ``All In All Out" (AIAO) antiferromagnetism \cite{Mola07,Mola09,Mola09a}.  However, because the zero-frequency QSI correlations may be masked by strong fluctuations and difficult to detect experimentally, it has been emphasized that magnetization measurements at mK temperatures are ideal to identify a QSI, since they probe the zero frequency (and zero wave vector) response. 

The possible existence of QSI in {\tto} has been addressed by  Baker et al. \cite{Baker11}, who infer the existence of magnetization plateaux from ac susceptibility data down to 25\,mK, as well as $\mu$SR measurements which exhibit peaks and kinks in the spin lattice relaxation rate consistent with the boundaries of the magnetization plateaux.  In another study, l'Hotel et al. \cite{Lhotel11} recently report extraction magnetometry and ac susceptibility down to 80\,mK claiming the absence of magnetization plateaux. However, data reported in the former study probe the response at \textit{finite} frequency, while the latter study does not extend to low enough temperatures to be conclusive. Moreover, extraction magnetometry suffers from the risk of tiny sample vibrations in the magnetic field, thereby changing the field history of the sample. Finally, both studies addressed a $\langle111\rangle$ direction only, not reporting the properties of other directions as a control experiment. Taken together, the nature of the spin liquid state in {\tto} and the proposal of QSI are hence unresolved.

In this Letter we address, to the best of our knowledge for the first time, the properties of {\tto} and the possible existence of QSI down to sufficiently low temperatures under conditions avoiding parasitic modifications of the field history. The magnetization of a {\tto} single-crystal was measured at TUM using a bespoke vibrating-coil magnetometer (VCM) for temperatures down to $\sim0.04\,{\rm K}$ and magnetic fields up to 5\,T \cite{Legl10a,Legl10b}. Our VCM represents a new development previously thought to be prohibitively difficult. As the main advantage of the VCM, the sample does not move during measurements (the detection system is completely decoupled from the dilution refrigerator). The coil-set was operated at 36.5\,Hz and the sample temperature was monitored with several RuO$_2$ and Speer sensors. Great care was taken to ensure good thermal anchoring (for details see \cite{Legl10b,Krey11}).

\begin{figure}
\includegraphics[width=0.4\textwidth]{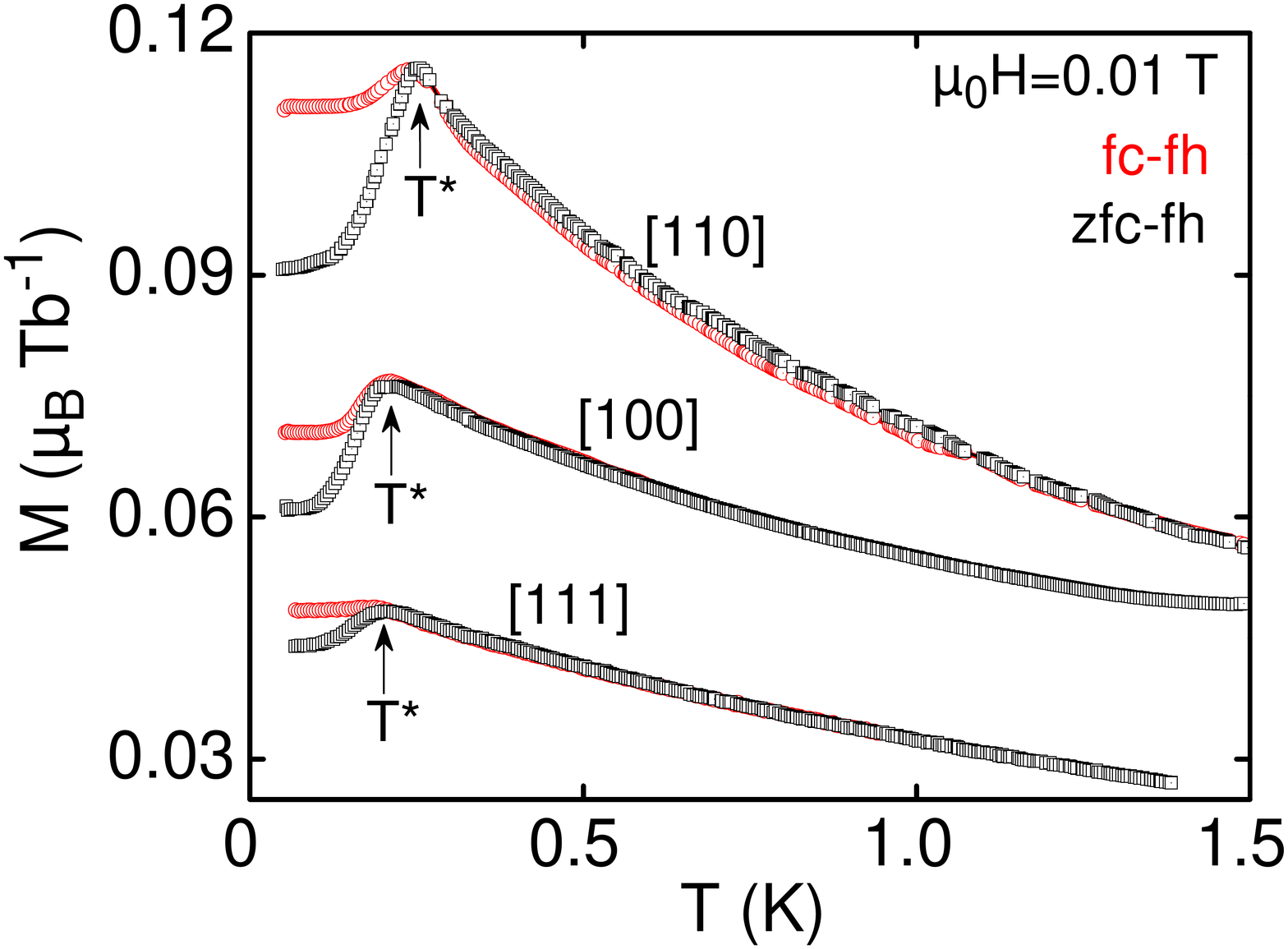}
\caption{(Colour online) Temperature dependence of the magnetization of {\tto} in a small applied magnetic field of 10\,mT. Below $T^*\sim 200\,{\rm mK}$ a distinct difference between data recorded under zero-field cooling (zfc) and field-cooling (fc) emerges, characteristic of a glassy magnetic state (all data were recorded while field-heating (fh) to minimise systematic errors). Curves are shifted for clarity as described in the text.}
\label{Fig_1}
\end{figure}

Magnetic field sweeps at up to 1\,T were recorded in a step mode, where the field was kept constant while recording the magnetization. Magnetization data up to 5\,T were recorded while sweeping the magnetic field continuously at 15 mT min$^{-1}$. For studies of a $\langle100\rangle$ and $\langle110\rangle$ direction the sample was mounted on a small copper wedge and the orientation confirmed with Laue x-ray diffraction. The empty sample holder was measured carefully and the background signal subtracted. The VCM data recorded below 1.5\,K was calibrated by means of a Ni standard measured in the same parameter range.  A comparison to the extrapolated magnetization of the {\tto} sample measured at 1.8\,K in a Quantum Design MPMS system enabled the background contribution to be isolated.  High temperature data were recorded in an Oxford Instruments vibrating sample magnetometer.

For our study a single crystal was grown at McMaster University by optical float-zoning. The feed and seed rods were prepared by annealing high-purity Tb$_4$O$_7$ and TiO$_2$ in air. The sample was then float-zoned in a high-purity Ar atmosphere of 4\,bar with a rate of 7\,mm\,hr$^{-1}$ \cite{Gard98}. Powder x-ray diffraction of a small piece of the single crystal confirmed that the ingot was phase-pure with the correct crystal structure. A single-crystalline disc ($7\times4.1\times0.8\,{\rm mm^3}$) was cut from the ingot for the magnetization measurements. The disc was oriented perpendicular to a $\langle111\rangle$ direction within $\sim 1^{\circ}$. The magnetic properties of this crystal as recorded above $\sim1.5\,{\rm K}$ were in excellent agreement with the literature. Moreover, none of our data displayed anomalies in ${\rm d}M/{\rm d}T$ and thus the magneto-caloric effect, probing spurious features in the specific heat seen in low-quality {\tto} \cite{Chap10}.

Demagnetizing fields were corrected for the $[111]$ orientation, approximating our sample as an ellipsoid with a demagnetising factor $N=0.34$. As we will show, the conclusions drawn from our data for $[100]$ and $[110]$ do not depend on the correction of demagnetizing fields, so only raw data as a function of applied field are shown. The internal fields were estimated to deviate by a few degrees from the $[100]$ and $[110]$ directions, because the plane of the disc was tilted with respect to the applied field. In addition, the demagnetization fields in the tilted sample may have been slightly inhomogeneous.

Figure \,\ref{Fig_1} illustrates the temperature dependence of the magnetization in an applied field of $10\,{\rm mT}$, where data for $[100]$ and $[110]$ have been shifted by $0.03\mu_{\rm B}\,{\rm Tb^{-1}}$ and $0.06\mu_{\rm B}\,{\rm Tb^{-1}}$, respectively for clarity (the uncertainty in the constant background is $\sim\pm 0.015\mu_{\rm B}\,{\rm Tb^{-1}}$). Prior to recording these data the superconducting magnet was demagnetized at $\sim2.3\,{\rm K}$ to remove any parasitic remanent fields. After zero-field cooling the field was increased at a rate of $1\,{\rm mT\,min}^{-1}$ to the setpoint of 10\,mT and data recorded while heating the sample continuously with $5\,{\rm mK\,min^{-1}}$ up to $\sim1.5\,{\rm K}$ (zfc-fh). Following this the sample was cooled down with the field unchanged and data recorded while heating at the same rate (fc-fh).

With decreasing temperature the magnetization increases with a positive curvature consistent with the paramagnetic properties at high temperatures. In all field directions, the curves display a cusp in the zfc-fh and fc-fh data.  We find slightly different values of $T^*\sim0.246\,{\rm K}$, $\sim0.212\,{\rm K}$ and $\sim0.204\,{\rm K}$ for $[110]$, $[100]$ and $[111]$, respectively, not reported previously. For $T>T^*$ the zfc-fh and fc-fh data agree essentially (the nature of the tiny difference for $[110]$ is most likely associated with a small drift of the detection system). 

The shape of the cusp, the absolute difference of zfc-fh and fc-fh data and the qualitative temperature dependence of the data provide strong evidence of the emergence of intrinsic magnetic glassiness below $T^*$ which is essentially isotropic. Note that magnetic ordering coexisting with persistent low temperature fluctuations has been reported in a variety of geometrically frustrated systems~\cite{lee,petrenko}. It is therefore perfectly consistent with the persistent muon spin relaxation observed in in {\tto} using $\mu$SR \cite{Gard99, Duns03}. The absolute size of the magnetization for $T>T^*$, being largest for [110] and smallest for [111], suggests that the tiny variation in $T^*$ and the difference between fc-fh and zfc-fh, which are also largest for [110] and smallest for [111], originate from the magnetic anisotropy. However, in comparison to simple cubic systems the anisotropy is unusual, as [110] cannot be a soft direction. 

\begin{figure}
\includegraphics[width=0.45\textwidth]{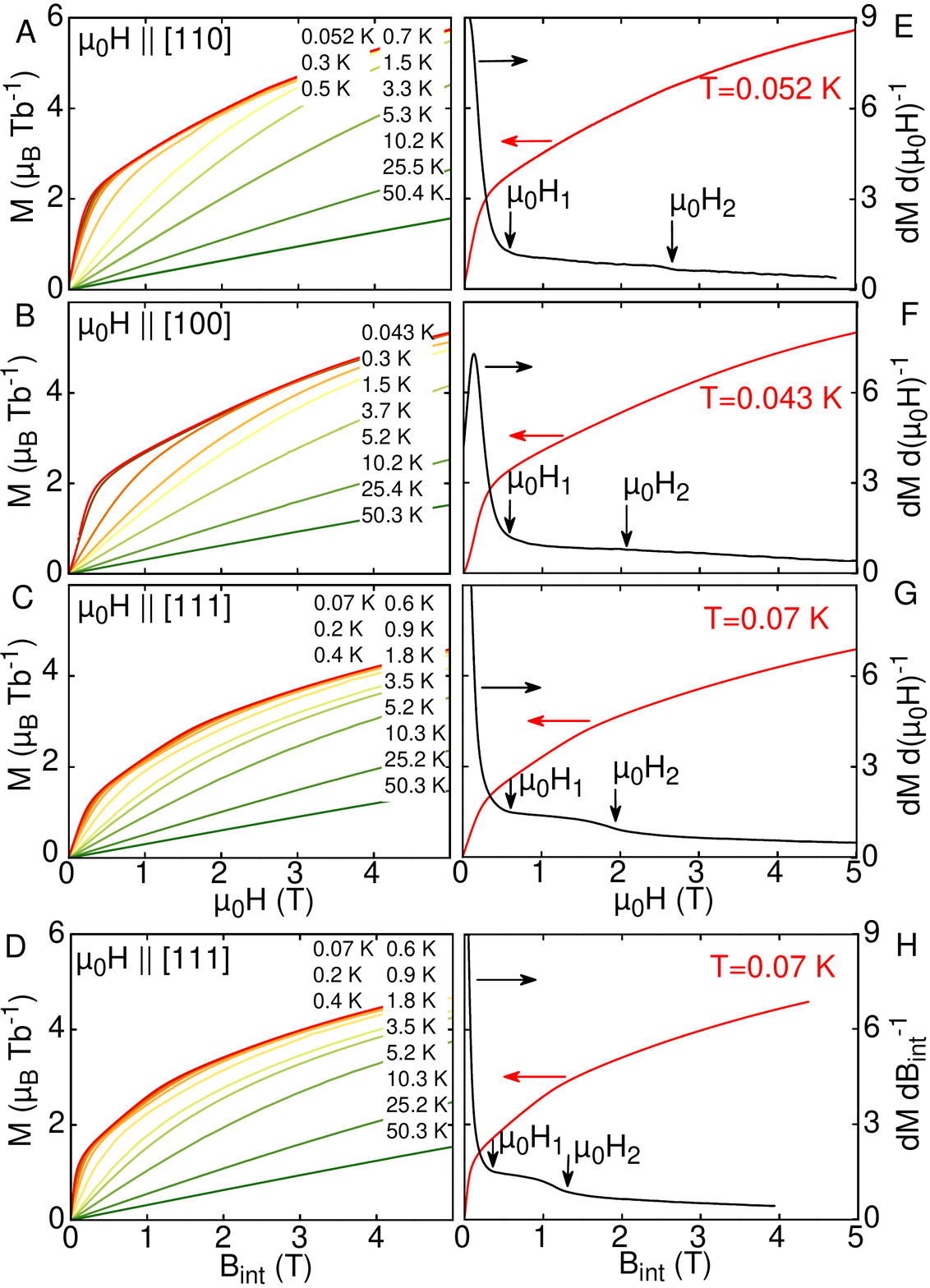}
\caption{(Color online) Magnetization of {\tto} for temperatures in the range $\sim0.043\,{\rm K}$ to $\sim50\,{\rm K}$ under magnetic fields up to 5\,T. Data below 1\,K were recorded after zfc. Panels (A), (B) and (C) display data as a function of the applied field along $[110]$, $[100]$ and $[111]$. Panel (D) shows the data of panel (C) as a function of internal field. Panels (E), (F), (G) and (H) show the numerical derivative of the experimental data recorded at the lowest temperatures in order to illustrate the definition of the characteristic fields $\mu_0\,H_1$ and $\mu_0\,H_2$.}
\label{Fig_2}
\end{figure}

Presented in Figs.\,\ref{Fig_2}\,(A), (B), (C) and (D) are magnetization data in the range from $\sim0.043\,{\rm K}$ to $\sim50\, {\rm K}$ under magnetic fields up to 5\,T for $[110]$, $[100]$ and $[111]$, respectively. Data below 1\,K were recorded after zero-field cooling. For the field and temperature scale shown in Fig.\,\ref{Fig_2} field-cooled data are identical. Note that panel (D) shows the data of panel (C) as a function of calculated internal field. Apart from a shift of the characteristic features in the magnetization towards lower fields, demagnetizing fields do not affect the conclusions of our study. For the temperature and field scale shown here data are qualitatively rather similar. The linear magnetic field dependence at $\sim50\,{\rm K}$ becomes highly non-linear at the lowest temperatures studied.  It remains unsaturated at $5\,{\rm T}$, even though the magnetization reaches a large value between 5 and $6\,{\rm \mu_B\,Tb^{-1}}$. 

On the scale shown in Fig.\,\ref{Fig_2} we find no evidence for a magnetization plateau, predicted for QSI. This contrasts with classical spin ice systems, where the magnetization plateaux are a  prominent feature strictly along $\langle111\rangle$ on the same scale. Moreover, we observe some fine-structure in {\dmdh} not reported before (Figs.\,\ref{Fig_2}\,(E), (F), (G) and (H)).  Namely, for the lowest temperature the calculated first derivatives allow us to define two cross-over scales $\mu_0\,H_1$ and $\mu_0\,H_2$.  The field $\mu_0\,H_1$ marks the end of the initial rise of the magnetization associated with the initial drop of {\dmdh}, while $\mu_0\,H_2>\mu_0\,H_1$ marks a faint drop of {\dmdh}. Thus the magnetization for $\mu_0\,H_1<B<\mu_0\,H_2$ cannot correspond to a plateau (or remnants thereof), since the slope \textit{decreases} when exceeding $\mu_0\,H_2$ without a point of inflection at $\mu_0\,H_2$. In addition, $\mu_0\,H_1$ and $\mu_0\,H_2$ exist for all directions and are not specific for $[111]$. 

Based on the isotropic behaviour observed, we speculate that microscopic features of $\mu_0\,H_1$ and $\mu_0\,H_2$ may have been seen in time-of-flight neutron scattering for field parallel $\langle110\rangle$, where the magnetic diffuse scattering, condenses into a new Bragg peak around $\mu_0\,H_1$ consistent with a polarized paramagnet \cite{Rule06}. A magnetically ordered phase, which supports spin wave excitations is induced around $\mu_0\,H_2$, consistent with cross-over phenomena around in the specific heat and ac susceptibility~\cite{Rule06}. Above $\mu_0\,H_2$ magnetization plateaux characteristic of spin ice are therefore clearly no longer expected. 

\begin{figure}
\includegraphics[width=0.45\textwidth]{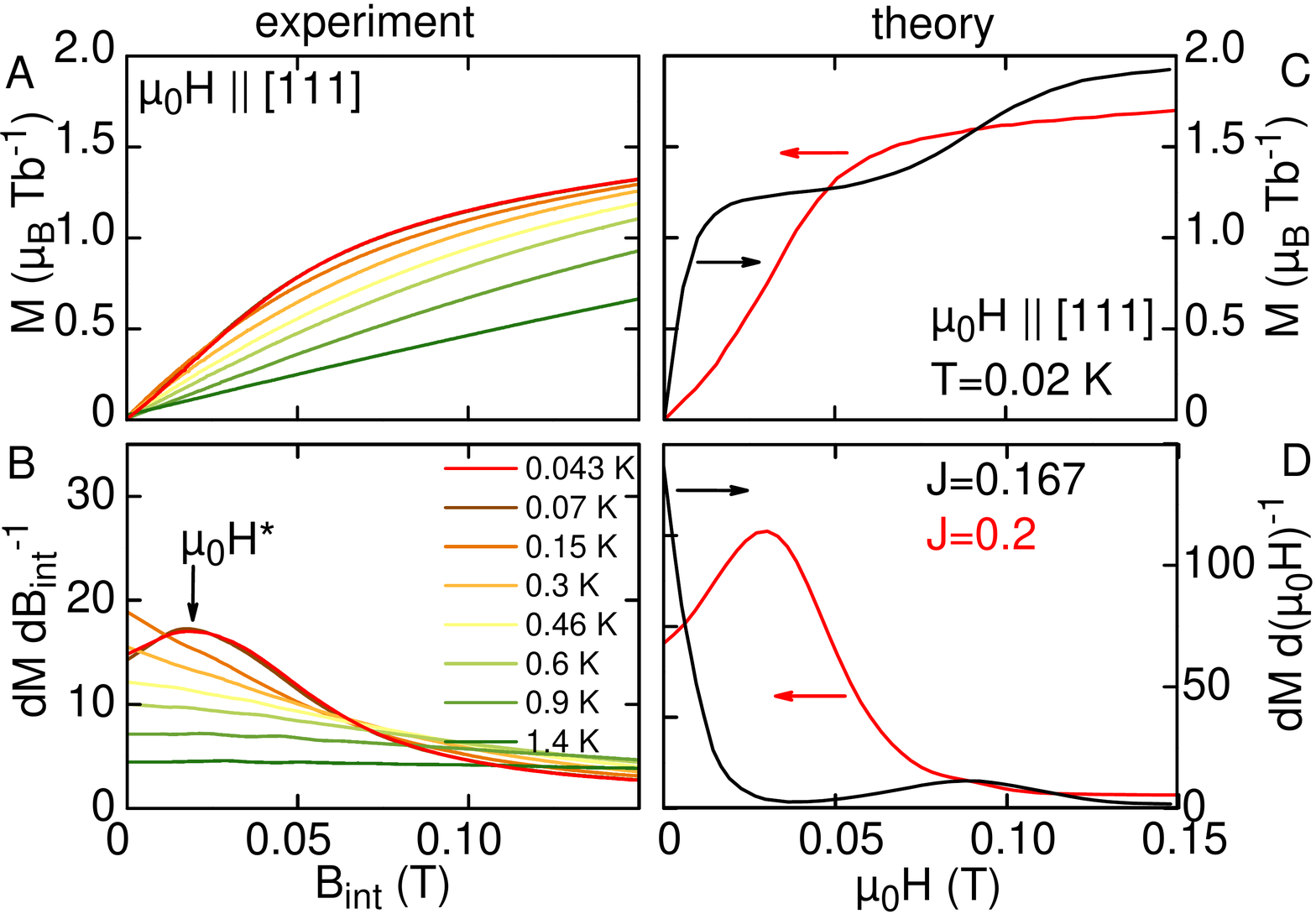}
\caption{(Color online) Experimental and theoretical low-field magnetization of {\tto} for the $[111]$ axis in the parameter range of the predicted magnetization plateau of quantum spin ice. Only zero-field cooled data are shown. (A) Magnetic field dependence of the magnetization of {\tto} in small  fields at various temperatures. (B) Numerical derivative of the data shown in panel (A). (C) Theoretically predicted magnetization for a quantum spin ice state in {\tto} as reported in Ref.\,\cite{Mola09a}. (D) Derivative of the theoretical data shown in panel (C).}
\label{Fig_3}
\end{figure}

Shown in Fig.\,\ref{Fig_3}\,(A) are typical zero-field cooled magnetization data for the $[111]$ direction, where we find no evidence of magnetization plateaux either. Qualitatively field-cooled data are the same as shown in the supplementary material. The absence of magnetization plateaux is most evident in {\dmdbi} calculated from the data shown in Fig.\,\ref{Fig_3}\,(B). This plot does not display a point of inflection of $M(B_{\rm int})$. Instead, {\dmdbi} has a broad maximum only. Data for $[110]$ and $[100]$ shown in the supplement are similar to $[111]$. Our data are thereby consistent with Ref.\,\cite{Lhotel11}, but extend a factor of two lower in temperature, well into the proposed QSI regime.

For comparison with the experimental data we reproduce in Fig.\,\ref{Fig_3}\,(C) and (D) theoretical calculations of the magnetization and their first derivatives at 20\,mK for $J=0.167$ (``quantum spin ice" (QSI)) and $J=0.2$ (``all-in/ all-out" (AIAO))~\cite{Mola09a}. The most important qualitative difference between the QSI and AIAO concerns the marked change in the slope of {\dmdh} at low field ($<0.05$ T) from negative to positive for QSI and AIAO structures, respectively.   Notice that experimentally {\dmdbi} at 43 and 70\,mK has a positive slope at low field and a maximum at 0.03\,T.  This observation is independent of the field history as described above.

However, the calculated AIAO spin state is based on a crystal field splitting 1/$\Delta $ which is reduced by a factor of two as compared with experiment and an exchange coupling, $J=0.2$, that is larger than that inferred experimentally from the AIAO spin structure in large magnetic fields \cite{Mire07}. Unfortunately we cannot offer an explanation for the discrepancy of the crystal field splitting. As for the exchange coupling it is important to note that the experimental value was determined in large magnetic fields  and may therefore differ from low fields \cite{Mire07}. Moreover, as shown in Ref.\,\cite{Mola09} for $J_{\rm ex}=1/6\,{\rm K}$ the predicted ground state depends on the model and its implementation~\cite{Mola09}, where the dipolar spin ice model and a cubic unit cell model with Ewald summed dipole-dipole interactions and crystal electric fields predict AIAO antiferromagnetism and long-range spin ice with ordering wave vector $\vec{q}=(000)$, LRSI$_{000}$, respectively.  Note however that the phase diagram is sensitive to the model used~\cite{mcclarty10}.  In particular, the model of Refs. [6-8] completely ignores the possibility of anisotropic exchange and higher multipolar couplings.  Because {\tto} does not develop spontaneous long-range magnetic order, it seems natural to conclude that the spin liquid in {\tto} is dominated by strong fluctuations at the border of AIAO-antiferromagnetism. This may be referred to as incipient AIAO-antiferromagnetism.

\begin{figure}
\includegraphics[width=0.3\textwidth]{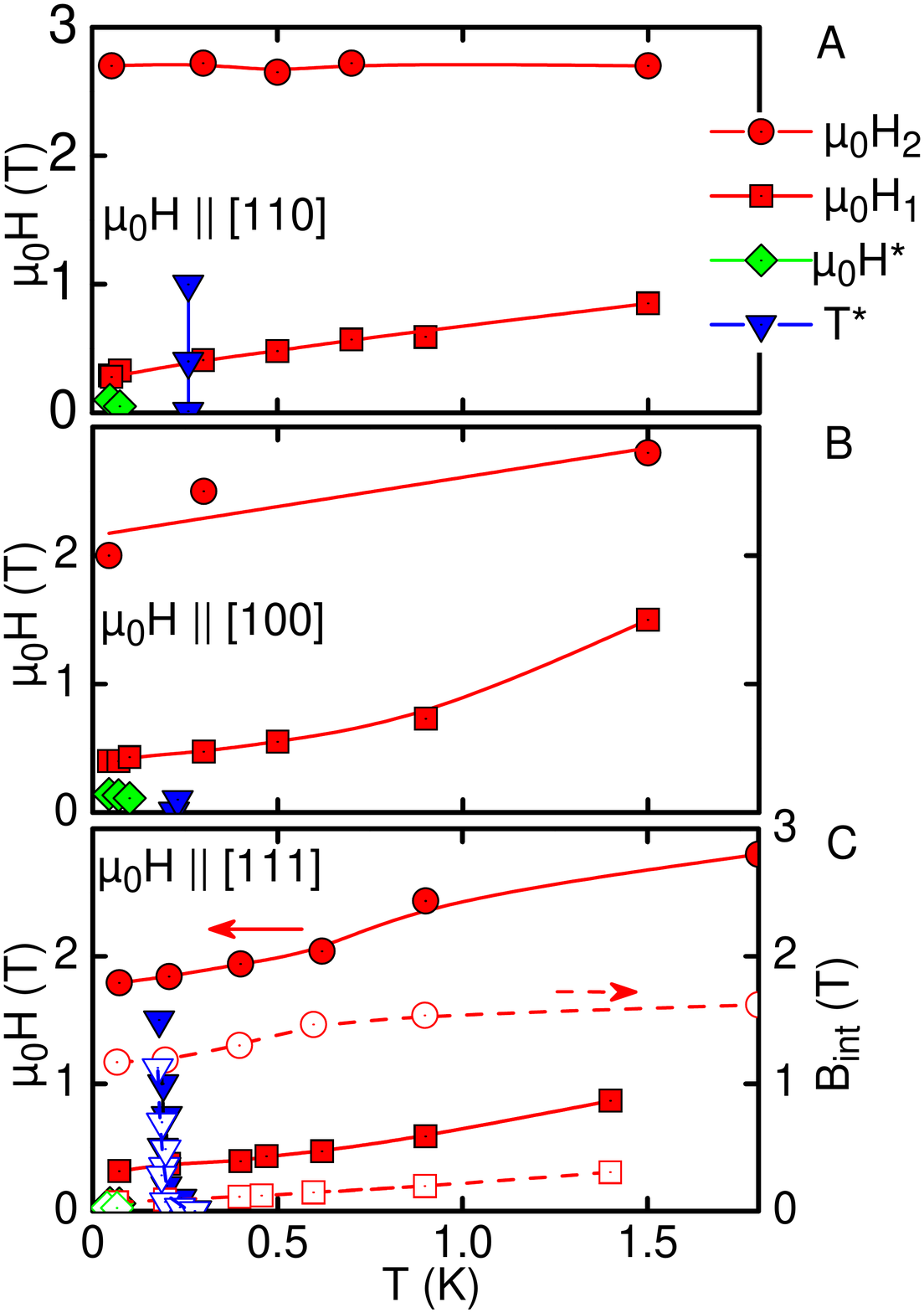}
\caption{(Color online) Magnetic phase diagram of {\tto} for $[110]$, $[100]$ and $[111]$ as inferred from the magnetization. The cross-over fields $\mu_0\,H_1$ and $\mu_0\,H_2$ are larger the magnetically softer the direction. This may serve as a testing ground for the validity of theoretical models. Open symbols in panel (C) correspond to the phase diagram after correction of demagnetising effects.}
\label{Fig_4}
\end{figure}

The nearly isotropic magnetic phase diagrams inferred from our magnetization data for $[110]$, $[110]$ and $[111]$ shown in Fig.\,\ref{Fig_4} corroborate incipient AIAO antiferromagnetism. Note that the open symbols in panel (C) show the phase diagram when correcting demagnetising effects.  The expected corrections in panels (A) and (B) are smaller than for panel (C). As a function of magnetic field we find two cross-over scales $\mu_0\,H_1$ and $\mu_0\,H_2$, where neutron scattering suggests different changes of the underlying microscopic properties \cite{Rule06}. 

In conclusion, we find no evidence of magnetization plateaux in {\tto} expected of QSI and fluctuation-induced ferromagnetic interactions. Instead, our data are in semi-quantitative agreement with the theoretical predictions of AIAO-antiferromagnetism, suggesting that the spin liquid state in {\tto} may be viewed as an incipient AIAO antiferromagnet. The small remaining orientational dependence in $T^*$, $\mu_0\,H_1$ and $\mu_0\,H_2$ reported in this Letter, which corresponds with the magnitude of the magnetization and thus the magnetic anisotropy, no doubt provides an important clue in future studies as to what inhibits the formation of long-range AIAO antiferromagnetism in {\tto} and drives the spin liquid state microscopically.

We wish to thank A. Bauer, P. B\"oni, B. Gaulin, S. Mayr, R. Moessner, R. Ritz, A. Regnat and C. Franz for support and stimulating discussions. We also wish to thank M. Gingras for a critical reading of the manuscript and M. Opel for support with the MPMS measurements at the WMI. Financial support through DFG TRR80 and FOR960, the Canadian Institute for Advanced Research (CIfAR) and NSERC is gratefully acknowledged.

%\bibliography{Tb2Ti2O7}
%\end{document}

%%%%%%%%%%%%%%%%%%%%%%%%%%%%%%%%%%%%%%%

%%%%%%%%%%%%%%%%%%%%%%%%%%%%%%%%%%%%%%%

\section{Supplementary Material for:  Vibrating-coil magnetometry of the spin liquid properties of \tto}

\begin{figure}[b]
\includegraphics[width=0.5\textwidth]{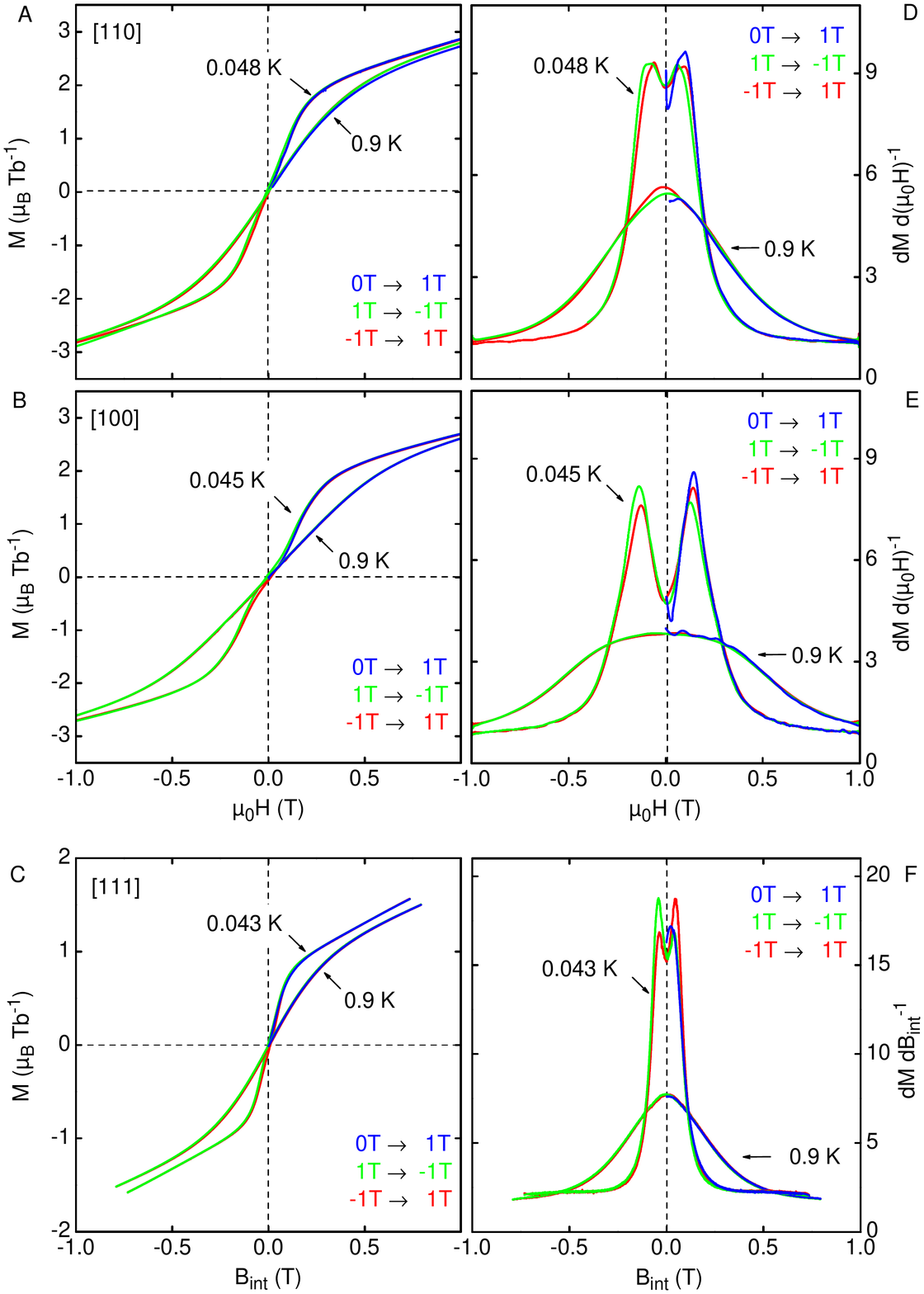}
\caption{(Color online).
Typical magnetization data of our {\tto} data comparing zero-field cooled with field cooled behaviour. Qualitatively no difference is observed regardless of orientation.}
\label{Fig_s1}
\end{figure}

Shown in Fig.\,\ref{Fig_s1} are low temperature magnetization data of our {\tto} sample.  For clarity only two temperatures are shown, one well below and the other well above $T^*$. For these measurements the VCM and sample were at first demagnetised at 2.3\,K and subsequently cooled to the desired temperature. Data were recorded in a sequence of three field sweeps at constant temperature:  (i) from $\mu_0\,H=0$ to $+1\,{\rm T}$, (ii) from $+1\,{\rm T}$ to $-1\,{\rm T}$, and (iii) from $-1\,{\rm T}$ to $+1\,{\rm T}$. Thus, the first sweep represents the zero-field cooled magnetization and the second and third sweep the field-ccoled magnetization. 

Data for field along $[110]$ and $[100]$ are shown in Fig.\,\ref{Fig_s1}\,(A) and (B) as a function of  applied magnetic field without correction for demagnetising effects. Data for $[111]$ are shown as a function of internal field in Fig.\,\ref{Fig_s1}\,(C) taking into account demagnetising fields. The fine structure of the slope of the magnetization is best revealed in the calculated first derivative of the magnetization. At the lowest temperatures measured the derivative displays a small maximum, regardless of sample orientation and field history in contrast to the prediction of a QSI.

%%%%%%%%%%%%%%%%%%%%%%%%%%%%%%%%%%%%%%%
\end{document}